\documentstyle[amssymb,floats,aps,prb,epsf]{revtex}
\epsfclipon
\begin{document}
\twocolumn[\hsize\textwidth\columnwidth\hsize\csname
@twocolumnfalse\endcsname

\draft
\draft
\title{Heat transport of electron-doped Cobaltates }
\author{Bin Liu, Ying Liang, and Shiping Feng}
\address{Department of Physics, Beijing Normal University,
Beijing 100875, China}
\author{Wei Yeu Chen}
\address{Department of Physics, Tamkang University, Tamsui 25137,
Taiwan}

\maketitle
\begin{abstract}
Within the $t$-$J$ model, the heat transport of electron-doped
cobaltates is studied based on the fermion-spin theory. It is
shown that the temperature dependent thermal conductivity is
characterized by the low temperature peak located at a finite
temperature. The thermal conductivity increases monotonously with
increasing temperature at low temperatures T $<$ 0.1$J$, and then
decreases with increasing temperature for higher temperatures T
$>$ 0.1$J$, in qualitative agreement with experimental result
observed from Na$_{x}$CoO$_{2}$ .

\end{abstract}

]
\bigskip

\narrowtext

The sodium cobalt oxide Na$_{x}$CoO$_{2}$ has become of
considerable interest in the past few years because of its
unconventional physical properties\cite{1}. The structure of
Na$_{x}$CoO$_{2}$ consists of triangular CoO$_{2}$ sheets
separated by layers of Na$^{+}$ ions\cite{1,2}. This structure is
similar to cuprates in the sense that they also have a layered
structure of the square lattice of the CuO$_{2}$ plane separated
by insulating layers\cite{3}. In the half-filling, the cobaltate
is a Mott insulator, then the system becomes a strange metal by
the electron doping\cite{1,2}. Moreover, superconductivity in the
hydrated system Na$_{x}$CoO$_{2}\cdot 1.3$H$_{2}$O has been
observed in a narrow range of the electron doping
concentration\cite{4}, around the optimal doping $x\approx0.3$.
Although the ferromagnetic correlation is present in
Na$_{x}$CoO$_{2}$ for the large electron doping concentration
($x\approx 0.7$) \cite{5}, the antiferromagnetic (AF) short-range
spin correlation in Na$_{x}$CoO$_{2}$ and Na$_{x}$CoO$_{2}\cdot
y$H$_{2}$O in the low electron doping concentration ($x\approx
0.35$) has been observed from the nuclear quadrupolar resonance
and thermopower as well as other experimental measurements
\cite{6}. Therefore the unexpected finding of superconductivity in
electron-doped cobaltate has raised the hope that it may help
solve the unusual physics in doped cuprates.

The heat transport is one of the basic transport properties that
provides a wealth of useful information on carriers and phonons as
well as their scattering processes\cite{7}. In the conventional
metals, the thermal conductivity contains both contributions from
carriers and phonons\cite{7}. The phonon contribution to the
thermal conductivity is always present in the conventional metals,
while the magnitude of the carrier contribution depends on the
type of materials because it is directly proportional to the free
carrier density. However, the phonon contribution to the thermal
conductivity is strongly suppressed for doped cuprates on a square
lattice\cite{8}. Moreover, an unusual contribution to the thermal
conductivity of doped two-leg ladder materials has been
observed\cite{9}, and this unusual contribution may be due to an
energy transport via magnetic excitations\cite{9}. Recently, the
heat transport of the electron-doped cobaltate Na$_{x}$CoO$_{2}$
has been studied experimentally\cite{10}, where the behavior of
the temperature dependent thermal conductivity is similar to the
case of the doped two-leg ladder materials, and can not be
explained within the conventional models of phonon heat transport
based on phonon-defect scattering or conventional phonon-electron
scattering. Within the charge-spin separation fermion-spin
theory\cite{11}, the heat transports of doped cuprates on the
square lattice and doped two-leg ladder materials have been
studied\cite{12,13} based on the $t$-$J$ model, and the results
are in agreement with experiments\cite{8,9}. Since the strong
electron correlation is common for doped cuprates, two-leg ladder
materials, and cobaltates, then these systems may have the same
scattering mechanism that leads to the unusual heat transport. In
this paper, we apply the charge-spin separation fermion-spin
approach to study the heat transport of electron-doped cobaltates.
Our results show that the thermal conductivity of the
electron-doped cobaltate Na$_{x}$CoO$_{2}$ is characterized by the
low temperature peak located at a finite temperature. Our results
also show that although both dressed charge carriers and spin are
responsible for the heat transport of electron-doped cobaltates,
the contribution from the dressed spins are dominant.

As in the doped cuprates, the two-dimensional CoO$_{2}$ plane
dominates the unconventional physics of electron-doped cobaltates.
It has been argued that the essential physics of the doped
CoO$_{2}$ plane can be described by the $t$-$J$ model on a
triangular lattice \cite{14},
\begin{eqnarray}
H&=-&t\sum_{i\hat{\eta}\sigma}PC_{i\sigma}^{\dagger}
C_{i+\hat{\eta}\sigma}P^{\dagger}-\mu\sum_{i\sigma}P C_{i\sigma
}^{\dagger}C_{i\sigma }P^{\dagger}\nonumber\\
&+&J\sum_{i\hat{\eta}}{\bf S}_{i}\cdot{\bf S}_{i+\hat{\eta}},
\end{eqnarray}
where summation is over all site $i$, and for each $i$ over its
nearest-neighbor $\hat{\eta}$, $C^{\dagger}_{i\sigma}$
($C_{i\sigma}$) is the electron creation (annihilation) operator,
${\bf S}_{i}=C^{\dagger}_{i}{\bf \sigma}C_{i}/2$ is the spin
operator with ${\bf \sigma}=(\sigma_{x},\sigma_{y},\sigma_{z})$ as
the Pauli matrices, $\mu$ is the chemical potential, and the
projection operator $P$ removes zero occupancy, i.e.,
$\sum_{\sigma}C^{\dagger}_{i\sigma} C_{i\sigma}\geq 1$. In order
to use the charge-spin separation fermion-spin
transformation\cite{11}, the $t$-$J$ model (1) can be rewritten in
terms of a particle-hole transformation $C_{i\sigma}\rightarrow
f^{\dagger}_{i-\sigma}$ as\cite{15},
\begin{eqnarray}
H=t\sum_{i\hat{\eta}\sigma}f_{i\sigma}^{\dagger}
f_{i+\hat{\eta}\sigma}+\mu\sum_{i\sigma}f_{i\sigma }^{\dagger}
f_{i\sigma }+J\sum_{i\hat{\eta}}{\bf S}_{i} \cdot{\bf
S}_{i+\hat{\eta}},
\end{eqnarray}
then the original local constraint
$\sum_{\sigma}C^{\dagger}_{i\sigma} C_{i\sigma}\geq 1$ is
transferred as $\sum_{\sigma}f^{\dagger}_{i\sigma}f_{i\sigma}\leq
1$ to remove double occupancy, where $f^{\dagger}_{i\sigma}$
($f_{i\sigma}$) is the hole creation (annihilation) operator, and
${\bf S}_{i}=f^{\dagger}_{i} {\bf \sigma}f_{i}/2$ is the spin
operator in the hole representation. In this case, the hole
operators can be expressed as
$f_{i\uparrow}=a^{\dagger}_{i\uparrow}S^{-}_{i}$ and
$f_{i\downarrow}=a^{\dagger}_{i\downarrow}S^{+}_{i}$ in the
charge-spin separation fermion-spin representation\cite{11}, where
the spinful fermion operator $a_{i\sigma}=
e^{-i\Phi_{i\sigma}}a_{i}$ describes the charge degree of freedom
together with some effects of the spin configuration
rearrangements due to the presence of the doped charge carrier
itself (dressed charge carrier), while the spin operator $S_{i}$
describes the spin degree of freedom (dressed spin), then the
local constraint, $\sum_{\sigma}
f^{\dagger}_{i\sigma}f_{i\sigma}=S^{+}_{i}a_{i\uparrow}
a^{\dagger}_{i\uparrow}S^{-}_{i}+S^{-}_{i}a_{i\downarrow}
a^{\dagger}_{i\downarrow}S^{+}_{i}=a_{i}a^{\dagger}_{i}(S^{+}_{i}
S^{-}_{i}+S^{-}_{i}S^{+}_{i})=1-a^{\dagger}_{i}a_{i}\leq 1$, is
satisfied in analytical calculations, and the double dressed
charge carrier occupancy,
$a^{\dagger}_{i\sigma}a^{\dagger}_{i-\sigma}=
e^{i\Phi_{i\sigma}}a^{\dagger}_{i}a^{\dagger}_{i}
e^{i\Phi_{i-\sigma}}=0$ and $a_{i\sigma}a_{i-\sigma}=
e^{-i\Phi_{i\sigma}}a_{i}a_{i}e^{-i\Phi_{i-\sigma}}=0$, are ruled
out automatically. It has been shown\cite{11} that these dressed
charge carrier and spin are gauge invariant, and in this sense,
they are real and can be interpreted as the physical excitations.
Although in common sense $a_{i\sigma}$ is not a real spinful
fermion, it behaves like a spinful fermion. In this charge-spin
separation fermion-spin representation, the low-energy behavior of
the $t$-$J$ model (2) can be expressed as,
\begin{eqnarray}
H&=&t\sum_{i\hat{\eta}}(a_{i\uparrow}S^{+}_{i}
a^{\dagger}_{i+\hat{\eta}\uparrow}S^{-}_{i+\hat{\eta}}+
a_{i\downarrow}S^{-}_{i}a^{\dagger}_{i+\hat{\eta}\downarrow}
S^{+}_{i+\hat{\eta}})\nonumber\\
&-&\mu\sum_{i\sigma}a^{\dagger}_{i\sigma} a_{i\sigma}+J_{{\rm
eff}}\sum_{i\hat{\eta}}{\bf S}_{i}\cdot {\bf S}_{i+\hat{\eta}},
\end{eqnarray}
with $J_{{\rm eff}}=(1-\delta)^{2}J$, and $\delta=\langle
a^{\dagger}_{i\sigma}a_{i\sigma}\rangle=\langle
a^{\dagger}_{i}a_{i}\rangle$ is the electron doping concentration.

According to the linear response theory\cite{16}, the thermal
conductivity can be obtained as,
\begin{eqnarray}
\kappa(\omega,T)=-{1\over T}{{\rm Im}\Pi_{Q} (\omega,T)\over
\omega},
\end{eqnarray}
with $\Pi_{Q}(\omega,T)$ is the heat current-current correlation
function, and is defined as,
\begin{eqnarray}
\Pi_{Q}(\tau-\tau')=-\langle T_{\tau}j_{Q}(\tau)j_{Q}(\tau')
\rangle,
\end{eqnarray}
where $\tau$ and $\tau'$ are the imaginary times, $T_{\tau}$ is
the $\tau$ order operator, while the heat current density $j_{Q}$
is obtained within the Hamiltonian (3) by using Heisenberg's
equation of motion as \cite{16},
\begin{mathletters}
\begin{eqnarray}
j_{Q}&=&i\sum_{i,j}{\bf R}_{i}[H_{i},H_{j}]=j^{(f)}_{Q}+
j^{(s)}_{Q}, \\
j^{(f)}_{Q}&=&i(\chi t)^{2}\sum_{i\hat{\eta}\hat{\eta}'\sigma}
\hat{\eta}a_{i+\hat{\eta}'\sigma}^{\dagger}a_{i+\hat{\eta}\sigma}-i\mu\chi
t\sum_{i\hat{\eta}\sigma}\hat{\eta}
a_{i+\hat{\eta}\sigma}^{\dagger}a_{i\sigma},\\
j^{(s)}_{Q}&=&i{1\over 2}(\epsilon J_{\rm{eff}})^{2}
\sum_{i\hat{\eta}\hat{\eta}'}(\hat{\eta}-\hat{\eta}')[S^{+}_{i}
S^{-}_{i-\hat{\eta}+\hat{\eta}'}S^{z}_{i-\hat{\eta}}\nonumber \\
&+&S^{-}_{i-\hat{\eta}+\hat{\eta}'}S^{+}_{i}S^{z}_{i-\hat{\eta}}]+i\epsilon
J^{2}_{\rm{eff}}\sum_{i\hat{\eta}\hat{\eta}'}
(\hat{\eta}-\hat{\eta}')[S^{+}_{i}S^{-}_{i+\hat{\eta}}
S^{z}_{i+\hat{\eta}'}\nonumber \\
&-&S^{+}_{i}S^{-}_{i-\hat{\eta}}S^{z}_{i-\hat{\eta}+\hat{\eta}'}],
\end{eqnarray}
\end{mathletters}
where ${\bf R}_{i}$ is lattice site, the spin correlation function
$\chi=\langle S_{i}^{+}S_{i+\hat{\eta}}^{-}\rangle$,
$\epsilon=1-2t\phi/J_{\rm eff}$, and the dressed charge carrier
particle-hole parameter $\phi=\langle
a^{\dagger}_{i\sigma}a_{i+\hat{\eta}\sigma}\rangle$. Within the
$t$-$J$ model, the thermal conductivity of hole-doped cuprates on
the square lattice has been obtained\cite{12}. Following their
discussions, the thermal conductivity of electron-doped cobaltates
on the triangular lattice can be obtained as,
\begin{mathletters}
\begin{eqnarray}
\kappa(\omega,T)&=&\kappa_{f}(\omega,T)+\kappa_{s}(\omega,T), \\
\kappa_{f}(\omega,T)&=&-{1\over 2N}\sum_{k\sigma}\Lambda_{f}^{2}
\gamma^{2}_{sk}\int^{\infty}_{-\infty}{d\omega'\over 2\pi}
A_{f\sigma}(k,\omega'+\omega) \nonumber \\
&\times& A_{f\sigma}(k,\omega'){n_{F}(\omega'+\omega)-n_{F}
(\omega') \over T\omega}, \\
\kappa_{s}(\omega,T)&=&-{1\over 2N}\sum_{k}\Lambda_{s}^{2}
\gamma^{2}_{sk}\int^{\infty}_{-\infty}{d\omega'\over 2\pi}
A_{s}(k,\omega'+\omega) \nonumber \\
&\times& A_{s}(k,\omega'){n_{B}(\omega'+\omega)-n_{B}(\omega')
\over T\omega},
\end{eqnarray}
\end{mathletters}
where $\Lambda_{f}=-Z\chi t(\mu+Z\chi t\gamma_{k})$, $\Lambda_{s}=
(ZJ_{{\rm eff}})^{2}\epsilon(2\epsilon\chi +2C-4\chi \gamma_{k})$,
$\gamma_{sk}^{2}=\{[{\rm sin}k_{x}+{\rm sin}(k_{x}/2) {\rm
cos}({\sqrt 3}k_{y}/2)]^{2}+3[{\rm sin}({\sqrt 3}k_{y}/2) {\rm
cos}(k_{x}/2)]^{2}\}/9$, $\gamma_{k}=[{\rm cos}(k_{x})+2{\rm
cos}(k_{x}/2){\rm cos}({\sqrt 3}k_{y}/2)]/3$,the spin spectral
function $A_{s}(k,\omega)=-2{\rm Im} D(k,\omega)$, the dressed
charge carrier spectral function $A_{f\sigma}(k,\omega)=-2{\rm
Im}g_{\sigma}(k,\omega)$, and $n_{F}(\omega)$ and $n_{B}(\omega)$
are the fermion and boson distribution functions, respectively,
while the full dressed charge carrier and spin Green's functions
are evaluated as,
\begin{mathletters}
\begin{eqnarray}
g^{-1}_{\sigma}(k,\omega)={1 \over
g^{(0)-1}_{\sigma}(k,\omega)-\Sigma_{f}(k,\omega)},\\
D^{-1}(k,\omega)={1\over
D^{(0)-1}(k,\omega)-\Sigma_{s}(k,\omega)},
\end{eqnarray}
\end{mathletters}
respectively, where the mean-field (MF) dressed charge carrier
Green's function $g^{(0)-1}_{\sigma}(k,\omega)=\omega-\xi_{k}$,
the MF spin Green's function
$D^{(0)-1}(k,\omega)=(\omega^{2}-\omega_{k}^{2})/B_{k}$, with
$B_{k}=\lambda[2\chi^{z}(\epsilon
\gamma_{k}-1)+\chi(\gamma_{k}-\epsilon)]$, $\lambda=2ZJ_{eff}$,
the spin correlation function $\chi^{z}=\langle
S_{i}^{z}S_{i+\hat{\eta}}^{z}\rangle$, and the MF dressed charge
carrier and spin excitation spectra are given by,
\begin{mathletters}
\begin{eqnarray}
\xi_{k}&=&-Zt\chi\gamma_{k}-\mu, \\
\omega^{2}_{k}&=&A_{1}(\gamma_{k})^{2}+A_{2}\gamma_{k}+A_{3},
\end{eqnarray}
\end{mathletters}
respectively, where
$A_{1}=\alpha\epsilon\lambda^{2}(\epsilon\chi^{z}+\chi/2)$,
$A_{2}=-\epsilon\lambda^{2}[\alpha(\chi^{z}+\epsilon\chi/2)+
(\alpha C^{z}+(1-\alpha)/(4Z)-\alpha\epsilon\chi/(2Z))+(\alpha
C+(1-\alpha)/(2Z)-\alpha\chi^{z}/2)/2]$, $A_{3}=\lambda^{2}
[\alpha C^{z}+(1-\alpha)/(4Z)-\alpha\epsilon\chi/(2Z)+\epsilon^{2}
(\alpha C+(1-\alpha)/(2Z)-\alpha\chi^{z}/2)/2]$, the spin
correlation functions $C=(1/Z^{2})
\sum_{\hat{\eta},\hat{\eta'}}\langle S_{i+\hat{\eta}}^{+}
S_{i+\hat{\eta'}}^{-}\rangle$,
$C^{z}=(1/Z^{2})\sum_{\hat{\eta},\hat{\eta'}}\langle
S_{i+\hat{\eta}}^{z}S_{i+\hat{\eta'}}^{z}\rangle$, while the
second-order dressed charge carrier and spin self-energy functions
are evaluated by the loop expansion to the second-order as,
\begin{mathletters}
\begin{eqnarray}
&~&\Sigma_{f}(k,\omega)={1\over 2}\left ({Zt\over N}\right)^2
\sum_{pp'}(\gamma^{2}_{p'+p+k}+\gamma^{2}_{p'-k})
{B_{p'}B_{p+p'}\over 4\omega_{p'}\omega_{p+p'}}\nonumber \\
&\times& \left ({F^{(h)}_{1}(k,p,p')\over \omega+\omega_{p+p'}-
\omega_{p'}-\xi_{p+k}}+{F^{(h)}_{2}(k,p,p')\over\omega+\omega_{p'}
-\omega_{p+p'}-\xi_{p+k}}\right. \nonumber \\
&+&\left.
{F^{(h)}_{3}(k,p,p')\over\omega+\omega_{p'}+\omega_{p+p'}
-\xi_{p+k}}-{F^{(h)}_{4}(k,p,p')\over\omega -\omega_{p+p'}-
\omega_{p'}-\xi_{p+k}}\right),\nonumber \\
\\
&~&\Sigma_{s}(k,\omega)=\left ({Zt\over N}\right )^{2}
\sum_{pp'}(\gamma^{2}_{p'+p+k}+\gamma^{2}_{p'-k})
{B_{k+p}\over 2\omega_{k+p}} \nonumber \\
&\times& \left
({F^{(s)}_{1}(k,p,p')\over\omega+\xi_{p+p'}-\xi_{p'}
-\omega_{k+p}}-{F^{(s)}_{2}(k,p,p')\over
\omega+\xi_{p+p'}-\xi_{p'}
+\omega_{k+p}}\right ), \nonumber \\
\end{eqnarray}
\end{mathletters}
respectively, where $F^{(h)}_{1}
(k,p,p')=n_{F}(\xi_{p+k})[n_{B}(\omega_{p'})-n_{B}(\omega_{p+p'})]
+n_{B}(\omega_{p+p'})[1+n_{B}(\omega_{p'})]$, $F^{(h)}_{2}(k,p,p')
=n_{F}(\xi_{p+k})[n_{B}(\omega_{p'+p})-n_{B}(\omega_{p'})]+n_{B}
(\omega_{p'})[1+n_{B}(\omega_{p'+p})]$, $F^{(h)}_{3}(k,p,p')=n_{F}
(\xi_{p+k})[1+n_{B}(\omega_{p+p'})+n_{B}(\omega_{p'})]+n_{B}
(\omega_{p'})n_{B}(\omega_{p+p'})$, $F^{(h)}_{4}(k,p,p')=n_{F}
(\xi_{p+k)}[1+n_{B}(\omega_{p+p'})+n_{B}(\omega_{p'})]-[1+n_{B}
(\omega_{p'})][1+n_{B}(\omega_{p+p'})]$,
$F^{(s)}_{1}(k,p,p')=n_{F}(\xi_{p+p'})[1-n_{F}(\xi_{p'})]-n_{B}
(\omega_{k+p})[n_{F}(\xi_{p'})-n_{F}(\xi_{p+p'})]$, and
$F^{(s)}_{2}
(k,p,p')=n_{F}(\xi_{p+p'})[1-n_{F}(\xi_{p'})]+[1+n_{B}
(\omega_{k+p})][n_{F}(\xi_{p'})-n_{F}(\xi_{p+p'})]$. In order not
to violate the sum rule of the correlation function $\langle
S^{+}_{i}S^{-}_{i}\rangle=1/2$ in the case without AF long-range
order, the important decoupling parameter $\alpha$ has been
introduced in the MF calculation \cite{17}, which can be regarded
as the vertex correction, then all the above MF order parameters,
decoupling parameter $\alpha$, and chemical potential $\mu$ are
determined by the self-consistent calculation \cite{18}.

We are now ready to discuss the heat transport of electron-doped
cobaltates. The observable temperature dependence of thermal
conductivity $\kappa(T)$ in the experiments can be obtained from
Eq. (7a) as $\kappa(T)=\lim_{\omega\rightarrow
0}\kappa(\omega,T)$. We have performed the numerical calculation
for this thermal conductivity $\kappa(T)$, and the results of
$\kappa(T)$ at the electron doping concentration $x=0.31$ (solid
line), and $x=0.33$ (dotted line) for $t/J=2.5$ are shown in Fig.
1. For the comparison, the experimental result\cite{10} taken from
Na$_{3.1}$CoO$_{2}$ is also shown in Fig. 1(inset). Our results
show that the thermal conductivity $\kappa(T)$ increases
monotonously with increasing temperatures for the low-temperature
range $T< 0.1J$, and reaches at the peak position in the
temperature $T\approx 0.1J$, then decreases with increasing
temperatures for the higher temperature range $T> 0.1J$. Using a
reasonable estimation value of $J\approx 300$ to $500k$ in
Na$_{x}$CoO$_{2}$, the position of the peak is at $T\approx 0.1J
\approx 30\sim50k$. These results are in qualitative agreement
with the experimental data \cite{10}.
\begin{figure}[prb]
\epsfxsize=3.5in\centerline{\epsffile{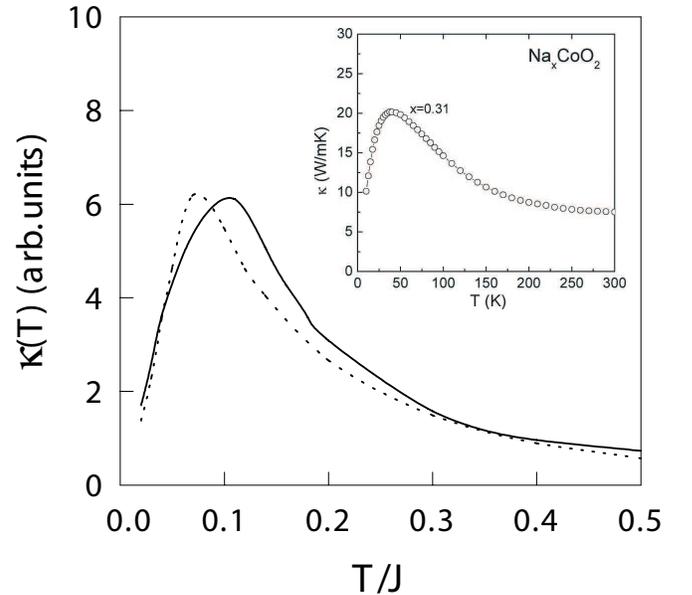}} \caption{The
thermal conductivity $\kappa(T)$ as a function of temperature at
the electron doping concentration x = 0.31 (solid line) and x =
0.33 (dotted line) with $t/J=2.5$.}
\end{figure}
During the above calculation, we have found as in doped
cuprates\cite{12} that although both dressed charge carriers and
spin are responsible for the thermal conductivity $\kappa(T)$, the
contribution from the dressed spins is much larger than these from
the dressed charge carriers, i.e., $\kappa_{s}(T)\gg
\kappa_{f}(T)$ in electron-doped cobaltates, and therefore the
thermal conductivity of electron-doped cobaltates is mainly
determined by its dressed spin part $\kappa_{s}(T)$. In this case,
the physical interpretation of the present results is the same as
in the case of doped cuprates\cite{12}, i.e., the observed unusual
thermal conductivity of electron-doped cobaltates is closely
related to the incommensurate spin response in the
systems\cite{19,20}. Since $\kappa_{s}(\omega,T)$ in Eq. (7c) is
obtained in terms of the dressed spin Green's function
$D(k,\omega)$, while the dynamical spin structure factor of doped
Mott insulators on the triangular lattice has been obtained from
the dressed spin Green's function as\cite{19,20},
\begin{eqnarray}
S(k,\omega)&=&-2[1+n_{B}(\omega)]{\rm Im}D(k,\omega)\nonumber \\
&=&-2[1+n_{B}(\omega)]B_{k}{\rm Im}\Sigma_{s}(k, \omega)\over
[\omega^{2}-\omega^{2}_{k}-{\rm Re} \Sigma_{s}
(k,\omega)]^{2}+[{\rm Im} \Sigma_{s}(k,\omega)]^{2},
\end{eqnarray}
where ${\rm Im}\Sigma_{s}(k,\omega)$ and ${\rm Re}
\Sigma_{s}(k,\omega)$ are corresponding imaginary part and real
part of the dressed spin self-energy function
$\Sigma_{s}(k,\omega)$ in Eq. (10b). As we have shown in detail in
Ref. [19,20], the dynamical spin structure factor (11) has a
well-defined resonance character. $S(k,\omega)$ exhibits a peak
when the incoming neutron energy $\omega$ is equal to the
renormalized spin excitation $E^{2}_{k}=\omega^2_{k}+B_{k}{\rm Re}
\Sigma_{s}(k,E_{k})$ for certain critical wave vectors
$k_{\delta}$ (positions of the incommensurate peaks), then the
height of these peaks is determined by the imaginary part of the
dressed spin self-energy function ${\rm Im} \Sigma_{s}(k_{\delta},
\omega)$. Since the time scale of this dynamical incommensurate
correlation is comparable to that of lattice vibrations as in the
case of the square lattice\cite{12}, then this dynamical spin
modulations dominate the heat transport of electron-doped
cobaltates, i.e., energy transport via magnetic excitations
dominates the thermal conductivity. On the other hand, the
dynamical spin response is doping dependent, this leads to the
thermal conductivity of electron-doped cobaltates also is doping
dependent.

In summary, we have studied the heat transport of electron-doped
cobaltates within the $t$-$J$ model. Our results show that the
thermal conductivity of electron-doped cobaltates is characterized
by the low temperature peak located at a finite temperature. The
thermal conductivity increases monotonously with increasing
temperatures at low temperature range $T< 0.1J$, and decreases
with increasing temperatures at higher temperature range $T>
0.1J$. Our results are in qualitative agreement with the
experimental data of Na$_{x}$CoO$_{2}$.

\acknowledgments One of us (BL) would like to thank Dr. Y.Y. Wang
for providing the experimental result of Na$_{x}$CoO$_{2}$. This
work was supported by the National Natural Science Foundation of
China under Grant Nos. 10404001 and  90403005, and the National
Science Council.

\end{document}